\documentclass[12pt,twoside]{article} 

\usepackage{graphicx}
\usepackage{balance}  
\usepackage{graphicx}
\usepackage{amsmath}
\usepackage[normalem]{ulem}
\usepackage{algorithmic}
\usepackage{algorithm}
\usepackage{multirow}
\usepackage{bu_ece_report}

\begin{document}

\buecedefinitions%
        {Characterization and Architectural Implications of Big Data Workloads}
        {}
        {LeiWang, Jianfeng Zhan, ZhenJia, RuiHan}
        {Beijing, China}
        {June.\ 25, 2015}
        {2015-6} 

\buecereporttitlepage



\buecereportsummary{Big data areas are expanding in a fast way in terms of increasing workloads and runtime systems, and this situation imposes a serious challenge to workload characterization, which is the foundation of innovative system and architecture design.  The previous major efforts on big data benchmarking either propose a comprehensive but a large amount of workloads, or only select a few workloads according to so-called popularity, which may lead to partial or even biased observations. In this paper, on the basis of a comprehensive big data benchmark suite---BigDataBench, we reduced 77 workloads to 17 representative workloads from a micro-architectural perspective. On a typical state-of-practice platform---Intel Xeon E5645, we compare the representative big data workloads with SPECINT, SPECCFP, PARSEC, CloudSuite and HPCC. After a comprehensive workload characterization, we have the following observations. First, the big data workloads are data movement dominated computing with more branch operations, taking up to 92\% percentage in terms of instruction mix, which places them in a different class from Desktop (SPEC CPU2006), CMP (PARSEC), HPC (HPCC) workloads. Second,  corroborating the previous work, Hadoop and Spark based big data workloads have higher front-end stalls.  Comparing with the traditional workloads i. e. PARSEC, the big data workloads have larger instructions footprint. But we also note that, in addition to varied instruction-level parallelism, there are significant disparities of front-end efficiencies among different big data workloads.  Third,  we found complex  software stacks that fail to use state-of-practise processors efficiently are one of the main factors leading to high front-end stalls. For the same workloads, the L1I cache miss rates have one order of magnitude differences among diverse implementations with different software stacks.}




\section{Introduction}  
A consensus on big data system architecture is based on a shared-nothing hardware design~\cite{stonebraker1986case, dewitt1992parallel} in which nodes communicate with one another only by sending messages via an interconnection network. Driven by cost-effectiveness,  scale-out solutions, which add more nodes to a system, are widely adopted to process an exploding amount of data. Big data are partitioned  across nodes, allowing multiple nodes to process large data in parallel, and this partitioned data and execution gives \emph{partitioned parallelism}~\cite{dewitt1992parallel}.  In nature, big data workloads are scale-out workloads mentioned ~\cite{ferdman-asplos-clearing-the-clouds}.

Previous system and architecture work shows that different software stacks, e.g., MapReduce or Spark, have significant impact on user-observed performance~\cite{zaharia2012resilient} and micro-architectural characteristics~\cite{jia_bigDataBench_subset}. In
addition to comprehensive workloads~\cite{hpca}, different software stacks should thus be included in the benchmarks~\cite{jia_bigDataBench_subset}, but that aggravates \emph{both cognitive difficulty on workload characterization and  benchmarking cost} by multiplying the number of workloads, which are the foundations of innovative system and architecture design.

To date, the previous major efforts on big data benchmarking either propose a comprehensive but a large amount of workloads (e.g.  a recent comprehensive big data benchmark suite---BigDataBench~\cite{hpca}, available from~\cite{BigDataBenchhomepage}, includes 77 workloads with different implementations)  or only select a few workloads according to so-called popularity~\cite{ferdman-asplos-clearing-the-clouds}, which may result in partial or biased observations.

In this paper, we choose 45 metrics from micro-architecture aspects, including instruction mix, cache and TLB behaviors, branch execution, pipeline behaviors, off-core requests and snoop response, parallelism, and operation intensity for workload characterization. On the basis of a comprehensive big data benchmark suite---BigDataBench, we reduces 77 workloads to 17 representative ones.
This reduction not only guarantees the comprehensiveness of big data workloads, but also significantly decreases benchmarking costs and cognitive difficulty on workload characterization.  We release the \emph{w}orkload \emph{c}haracterization and \emph{r}eduction \emph{t}ool named \emph{WCRT} as an open-source project.

We compare the seventeen representative big data workloads with SPECINT, SPECCFP, PARSEC, HPCC, CloudSuite, and TPC-C on the system consisting of Intel Xeon E5645 processors. To investigate the impact of different software stacks, we also add six workloads implemented with MPI (the same workloads included in the representative big data workloads). After a comprehensive workload characterization, we have the following observations.

First, for the first time, we reveal that the big data workloads have more branch and integer instructions, which place them in a different class from desktop (SPEC CPU2006), CMP (PARSEC), HPC (HPCC) workloads. Though analyzing the instructions breakdown, we found that the big data workloads are data movement dominated computing with more branch operations, which takes up to 92\% percentage in terms of instruction mix.


Second, corroborating the previous work, the Hadoop and Spark based big data workloads have higher front-end stalls.  Comparing with the traditional workloads i. e. PARSEC, the big data workloads have larger instructions footprints. But we also note that, in addition to varied instruction-level parallelism, there are significant disparities of front-end efficiencies among different subclasses of big data workloads. From this angle, previous work such as CloudSuite only covers a part of the workloads included in BigDataBench.

Third, we found complex software stacks that fail to use state-of-practise processors efficiently are one of the main factors leading to high front-end stalls. For the same workloads, the L1I cache miss rates have one order of magnitude differences among diverse implementations with different software stacks, which are overlooked in the previous work~\cite{ferdman-asplos-clearing-the-clouds} ~\cite{jia2013characterizing} ~\cite{hpca}. For the MPI-version of the big data workloads, their L1I numbers are very close to the traditional benchmarks. In addition to innovative hardware design, we should pay great attention to co-design of software and hardware so as to use state-of-practise processors efficiently.

\section{Background}
\subsection{BigDataBench}
BigDataBench is an open-source comprehensive big data benchmark suite. The current version--BigDataBench 3.0--includes 77 workloads covering four types of applications (cloud OLTP, OLAP and interactive analytics, and offline analytics) and three popular Internet scenarios (search engine, social network and e-commerce).
These workloads cover both basic operations and state-of-art algorithms, and each operation/algorithm has multiple implementations built upon mainstream software stacks such as Hadoop and Spark.
In short, BigDataBench aims at providing comprehensive workloads in order to meet the needs of benchmark users from different research fields such as architecture, system, and networking.

\subsection{WCRT}
WCRT is a comprehensive workload characterization tool, which can subset the whole workload set by removing redundant ones to facilitate workload characterization and other architecture research. It can also collect, analyze, and visualize a large number of performance metrics.
WCRT consists of two main modules: profilers and a performance data analyzer. On each  node, a \emph{profiler} is deployed to characterize workloads running on it. The profiler collects performance metrics specified by users once a workload begins to run, and transfers the collected data to the performance data analyzer when the workload completes. The analyzer is deployed on a dedicated node that does not run other workloads. After collecting the performance data from all profilers, the analyzer processes them using statistical and visual functions. The statistical functions are used to normalize performance data and perform principle component analysis.

\section{Representative big data Workloads}\label{Workloads}
To reduce the cognitive difficulty of workload characterization and benchmarking cost, we use WCRT to reduce the number of workloads in benchmarking from the perspective of micro-architecture. As the input for the WCRT tool, we choose 45 micro architecture level metrics, covering the characteristics of instruction mix, cache behavior, translation look-aside buffer (TLB) behavior, branch execution, pipeline behavior, off-core requests and snoop responses, parallelism, and operation intensity. Due to limited space, we give the details of these 45 metrics on our web page which is available from ~\cite{BigDataBenchhomepage}. Then we normalize these metric values to a Gaussian distribution and use Principle Component Analysis (PCA) to reduce the dimensions. Finally we use K-Means to cluster the 77 workloads, and there are 17 clusters in the final results.

\subsection{Original data sets of representative workloads}
There are seven data sets for representative workloads. As shown in Table \ref{data_sets}, these data have different types and sources and application domains. The original data set can be scaled by the BDGS provided by BigDataBench. The more details can be obtained from ~\cite{BigDataBenchhomepage}.
\begin{table}[H]
\caption{The summary of data sets and data generation tools.}\label{data_sets} \center \begin{tabular}{|p{1in}|p{1.5in}|p{1.5in}|p{1in}|}
\hline
No. &data sets & data set description &scalable data set\\ \hline
1 & Wikipedia Entries & 4,300,000 English articles& Text Generator of BDGS\\ \hline
2 & Amazon Movie Reviews & 7,911,684 reviews & Text Generator of BDGS\\ \hline
3 &Google Web Graph & 875713 nodes, 5105039 edges & Graph Generator of BDGS\\ \hline
4 &Facebook Social Network & 4039 nodes, 88234 edges &Graph Generator of BDGS\\ \hline
5 &E-commerce Transaction Data & Table 1: 4 columns, 38658 rows. Table
2: 6 columns, 242735 rows &Table Generator of BDGS\\ \hline
6 & ProfSearch Person Resum\'{e}s & 278956 resum\'{e}s&Table Generator of BDGS\\ \hline
7 & TPC-DS WebTable Data & 26 tables&TPC DSGen\\ \hline
\end{tabular} \end{table}

\subsection{Behaviors characteristics of representative workloads}
The representative workloads are implemented by different approaches. In Table~\ref{reduing workloads}, we give descriptions of each representative big data workload. We describe each workload from the perspective of system behaviors, data behaviors and application category.

\subsubsection{System Behaviors}
As the variations of big data workloads, we use the system behaviors to classify and characterize them. We choose CPU Usages, DISK IO behaviors and IO Bandwidth to analyze the system behaviors of big data workloads.
First, the CPU usages are described by \emph{CPU utilization} and \emph{IO Wait ratio}. \emph{CPU utilization} is defined as the percentage of time that the CPU executing at the system or user level, while \emph{I/O Wait ratio} is defined as the percentage of time that the CPU waiting for outstanding disk I/O requests. Second, DISK I/O performance is a key metric of big data workloads. We investigate the DISK I/O behavior with the \emph{average weighted Disk I/O time ratio}. \emph{Weighted disk I/O  time} is defined as the number of I/O in progress times the number of milliseconds spent doing I/O since the last update, and \emph{the average weighted Disk I/O time ratio} is the weighted Disk I/O  time divided by the running time of the workload.
Third, we choose the disk I/O bandwidth, network I/O bandwidth which can reflect the I/O throughput requirements of big data workloads. Based on the above metrics, we roughly classify the workloads into three category: \emph{CPU-intensive} workloads, which have high CPU utilizations, low average weighted Disk I/O time ratio or I/O Bandwidth;
\emph{I/O-intensive} workloads, which have high average weighted Disk I/O time ratio or I/O Bandwidth but low CPU utilizations;
and \emph{hybrid} workloads, whose behaviors are between CPU-intensive and IO-intensive workloads.
In this paper, the rule of classifying big data workloads is as follows: 1) For a workload, if the CPU utilization is larger than 85\%, we consider it CPU-Intensive; 2) For a workload, if the average weighted Disk I/O time ratio is larger than 10 or the I/O wait ratio is larger than 20\% and the CPU utilization is less than 60\%, we consider it I/O-Intensive; 3) other workloads excepting the CPU-intensive and I/O-intensive ones are considered as hybrid workloads.

\subsubsection{Data Behaviors}
For each workload, we characterize the data behaviors from perspective of data schema and data processing behaviors which measure the ratios of data input, output and intermediate data. For data schema, we will describe the data structure and semantic information of each workload. For data processing behaviors, we will describe the ratio of input and output and the intermediate data. We use \emph{larger}, \emph{less} and \emph{equal} to describe the data capacity changing. For example, when the ratio of the data output to the data input is larger than or equal to 0.9 and less than 1.1, we consider Output=Input; when the ratio of the data output to the data input is larger than or equal to 0.01 and less than 0.9, we consider Output\textless Input; when the ratio of the data output to the data input is less than 0.01, we consider Output\textless\textless Input; when the ratio of the data output to the data input is greater than or equal to 1.1, we consider Output\textgreater Input. The rule is inspired by Luo et al.~\cite{luo2012cloudrank}.

\subsubsection{Application category}
We consider three application categories: data analysis workloads, service workloads and interactive analysis workloads.
\begin{table*}
\tiny
\caption{Details of the representative big data workloads.} \newsavebox{\tablebox} \begin{lrbox}{\tablebox}
\label{reduing workloads}
\begin{tabular}{|p{0.2in}|p{0.8in}|p{1.5in}|p{0.8in}|p{0.8in}|p{0.8in}|p{0.8in}|}
\hline
\bf ID & \bf The name of representative big data workload (Abbr.) & \bf Description of the representative
workload & \bf Data Description & \bf category & \bf Data Processing Behaviors & \bf System behaviors\\ \hline

1 & HBase-Read (H-Read) (10)\footnotemark[1] & Basic operation of reading in HBase which is a popular non-relational, distributed database. & ProfSearch data set, each record is 1128 bytes K-V text file& service & Output=Input and no intermediate & IO-Intensive\\ \hline

2 & Hive-Difference (H-Difference) (9)\footnotemark[1]& Hive implementation of set difference, one of the five basic operator from relational algebra. & E-commerce Transaction data set, each record is 52 bytes K-V text file& interactive analysis & Output\textless Input and Intermediate\textless Input  &IO-Intensive\\ \hline

3 & Impala-SelectQuery (I-SelectQuery) (9)\footnotemark[1]& Impala implementation of select query to filter data, filter is one of the five basic operator from relational algebra. & E-commerce Transaction data set, each record is 52 bytes K-V text file  & interactive analysis & Output\textless Input and no Intermediate\textless\textless Input&IO-Intensive \\ \hline

4 & Hive-TPC-DS-Q3 (H-TPC-DS-query3) (9)\footnotemark[1]& Hive implementation of query 3 of TPC-DS, a popular decision support benchmark proposed by Transection Processing Performance Council, complex relational algebra. & TPC-DS Web data set, each record is 14 KB K-V text file & interactive analysis & Output=Input and no Intermediate &Hybrid\\ \hline

5 & Spark-WordCount (S-WordCount) (8)\footnotemark[1]& Spark implementation of word counting which counts the number of each word in the input
file. Counting is a a fundamental operation for big data statistics analytics. & Wikipedia data set, each record is 64KB K-V text file& data analysis & Output\textless\textless Input and Intermediate\textless Input& IO-Intensive\\ \hline

6 & Impala-OrderBy (I-OrderBy) (7)\footnotemark[1]& Impala implementation of sorting, a fundamental operation from relational algebra and extensively used in various scene. & E-commerce Transaction data set, each record is 52 bytes K-V text file & interactive analysis & Output=Input and Intermediate=Input&Hybrid\\ \hline

7 & Hadoop-Grep (H-Grep) (7)\footnotemark[1]& Searching plain text file for lines that match a regular expression by Hadoop MapReduce. Searching is another fundamental operation widely used. & Wikipedia data set, each record is 64KB K-V text file & data analysis &Output\textless\textless Input and Intermediate\textless\textless Input &CPU-Intensive\\ \hline

8 & Shark-TPC-DS-Q10 (S-TPC-DS-query10) (4)\footnotemark[1]& Shark implementation of query 10 of TPC-DS, complex relational algebra. & TPC-DS Web data set, each record is 14 KB K-V text file & interactive analysis &Output\textless\textless Input and no Intermediate &Hybrid\\ \hline

9 & Shark-Project (S-Project) (4)\footnotemark[1]& Shark implementation of project, one of the five basic operator from relational algebra. & E-commerce Transaction data set, each record is 52 bytes K-V text file & interactive analysis & Output\textless Input and no Intermediate &IO-Intensive\\ \hline

10 & Shark-OrderBy (S-OrderBy) (3)\footnotemark[1]& Shark implementation of sorting. &E-commerce Transaction data set, each record is 52 bytes K-V text file & interactive analysis &Output=Input and Intermediate=Input &IO-Intensive \\ \hline

11 & Spark-Kmeans (S-Kmeans) (1)\footnotemark[1]& Spark implementation of k-means which is a popular clustering algorithm in Discrete mathematics for partitioning n observations into k clusters .  &Facebook data set, each record is 94 bytes K-V text file & data analysis & Output=Input and Intermediate=Input&CPU-Intensive\\ \hline

12 & Shark-TPC-DS-Q8 (S-TPC-DS-query8) (1)\footnotemark[1]& Shark implementation of query 8 of TPC-DS.&TPC-DS Web data set, each record is 14 KB K-V text file & interactive analysis &Output\textless\textless Input and no Intermediate &Hybrid\\ \hline

13 & Spark-PageRank (S-PageRank) (1)\footnotemark[1]& Spark implementation of PageRank, which is a graph computing algorithm used by Google to score the importance of the
web page by counting the number and quality of links to the page.  &Google data set, each record is 6KB K-V text file & data analysis & Output\textgreater Input and Intermediate\textgreater Input&CPU-Intensive\\ \hline

14 & Spark-Grep (S-Grep) (1)\footnotemark[1]& Spark implementation of Grep.  &Wikipedia data set, each record is 64KB K-V text file&data analysis &Output\textless\textless Input and Intermediate\textless\textless Input &IO-Intensive \\ \hline

15 & Hadoop-WordCount (H-WordCount) (1)\footnotemark[1]& Hadoop implementation of WordCount.  &Wikipedia data set, each record is 64KB K-V text file & data analysis &Output\textless\textless Input and Intermediate\textless\textless Input &CPU-Intensive\\ \hline

16 & Hadoop-NaiveBayes (H-NaiveBayes) (1)\footnotemark[1]& Hadoop implementation of naive bayes which is a simple but widely used probabilistic classifier in statistical calculation.  &Amazon data set, each record is 52KB K-V text file & data analysis &Output\textless\textless Input and Intermediate\textless\textless Input &CPU-Intensive\\ \hline

17 & Spark-Sort (S-Sort) (1)\footnotemark[1]& Spark implementation of sorting. &Wikipedia data set, each record is 64KB K-V text file & data analysis & Output=Input and Intermediate=Input&Hybrid\\ \hline

\end{tabular}
\end{lrbox}
\scalebox{1.0}{\usebox{\tablebox}}
\footnotemark[1]{The number of workloads that the selected workloads can represent are given in parentheses.}
\end{table*}

\section{Experimental Configurations and Methodology}

This section presents experiment configurations and  methodology, respectively.

\subsection{Experiment Configurations}\label{system_configuration}
To obtain insights into the system and architecture for big data, we run a series of experiments using the seventeen representative workloads.

Jia et al.~\cite{jia_bigDataBench_subset} found that software stacks have a serious impact on  big data workloads in terms of micro-architectural characteristics.  Compared to traditional software stacks,
big data software stacks, e. g., Hadoop or Spark  usually have more complex structures, enabling  programmers to write less code to achieve
their intended goals. The upshot is two-fold. On one hand, the ratio of system
software and middleware instructions executed compared to
user applications instructions tends to be large, which makes
their impact on system behavior large, as well.  On the other hand, they have larger instruction footprint. For example, the previous work~\cite{ferdman-asplos-clearing-the-clouds}~\cite{jiacharacterization}~\cite{hpca} reported higher L1I cache miss rate for Hadoop and Spark workloads.  For comparison, in addition to BigDataBench subset, we also add MPI implementations of  six data analysis workloads, including Bayes, K-means, PageRank, Grep, WordCount and Sort.   The reason of choosing MPI is as follows: first, MPI can implement all of the operator primitives of Hadoop or Spark but with much sophisticated programming skill. Second, comparing with Hadoop and Spark, the MPI stacks is much  thinner.

For the same big data application, the scale of the system running big data applications is mainly decided by the size of the input data. For experiments in this paper, the input data is about 128GB, except \emph{PageRank} workload, which is measured in terms of the number of vertices. We deploy the big data workloads on the system with a matching scale---5 nodes. On our testbed, each node owns one Xeon E5645 processor equipped with 32 GB memory and 8 TB disk as listed in Table \ref{hwconfigeration}. In the rest of the experiments, hyperthreading is disabled on our testbed because  enabling these features makes it more complex
to measure and interpret performance data~\cite{levinthal2009performance}. The operating system is Centos 6.4 with Linux kernel 3.10.11. The Hadoop and JDK distribution is 1.0.2 and 1.6, respectively. The Spark distribution is 1.0.2. The HBase, Hive and MPICH2 distribution is 0.94.5, 0.9, 1.5, respectively. Table \ref{reduing workloads} shows the workload summary.

\begin{table}
\caption{Node configuration details of Xeon E5645}\label{hwconfigeration}
\center
\begin{lrbox}{\tablebox}
\begin{tabular}{|c|c|c|c|}
  \hline
  \multicolumn{2}{|c|}{\bf CPU type} & \multicolumn{2}{c|}{Intel \textregistered Xeon E5645} \\ \hline
  \multicolumn{2}{|c|}{\bf Number of cores}  &\multicolumn{2}{c|}{6 cores@2.40G} \\ \hline
  \hline
\bf L1 DCache &\bf L1 ICache &\bf L2 Cache &\bf L3 Cache \\ \hline
6 $\times$ 32 KB& 6 $\times$ 32 KB&6 $\times$ 256 KB& 12MB \\ \hline
\end{tabular}
\end{lrbox}
\scalebox{1.2}{\usebox{\tablebox}}
\end{table}

\subsection{Experiment Methodology}
Intel Xeon processors provide hardware performance counters to support micro-architecture level profiling.
We use Perf, a Linux profiling tool, to collect about dozens of events whose numbers and unit masks can be found in the Intel Developer's Manual. In addition, we access the proc file system to collect OS-level performance data. We collect performance data after a ramp up period, which is about 30 seconds.



\subsection{The Other Benchmarks Setup}
For \emph{SPEC CPU2006}, we ran the official applications with the first reference input, and separated the average results into two groups:
integer benchmarks (\emph{SPECINT}) and floating point benchmarks (\emph{SPECFP}). We have used \emph{HPCC} 1.4, which is a representative HPC benchmark suite, for the experiment. We ran all of the seven benchmarks in HPCC. \emph{PARSEC} is a benchmark suite composed of multi-threaded programs, and we deployed \emph{PARSEC} 3.0 Beta Release. We ran all the 12 benchmarks with native input data sets and used GCC 4.1.2 for compiling. \emph{CloudSuite} is a benchmark suite composed of Cloud scale-out workloads, and we deployed \emph{CloudSuite} 1.0 Release. We ran all the six benchmarks with input data that correspond with our benchmark data sets. TPC-C is an online transaction processing (OLTP) benchmarks, and our deployment  is tpcc-uva v1.2.

\section{Experiment Results and Observations} \label{E5645}
In this section, we report the micro-architecture behaviors through reporting instruction mix, pipeline efficiency and cache efficiency. Furthermore, to further understand the micro-architecture behaviors, we investigate the footprint of big data workloads and the software impacts for big data workloads. For better clarification, in addition to the average behaviors, we also report the behaviors of three subclasses of big data workload classified in Table 2. We compare the representative big data workloads with PARSEC, SPECINT, SPECFP, HPCC, CloudSuite and TPC-C.

Section ~\ref{Mix} introduces the instructions mix; section ~\ref{ILP} introduces instruction level parallelism; section ~\ref{cache_behavior} introduces the cache behaviors; section ~\ref{Footprint} introduces locality; section ~\ref{SoftwareStacks} introduces the software stacks impacts for big data workloads; section ~\ref{Summary} is the summary.

\subsection{Instruction Mix} \label{Mix}

\begin{figure}
\centering
\includegraphics[scale=0.8]{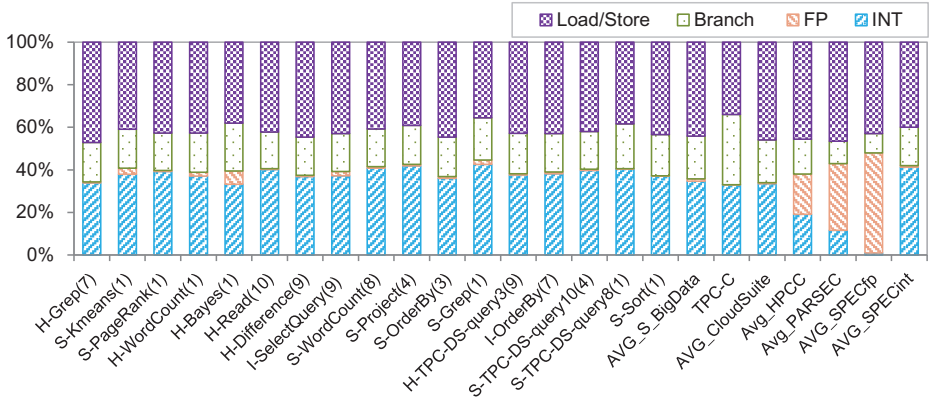}
\caption{The instruction breakdown of different workloads on X86 platforms.}
\label{figure:InsMix}
\end{figure}

In order to reveal instruction behaviors of big data workloads, we choose the instruction mix as the metric.
Figure~\ref{figure:InsMix} shows the retired instruction breakdown, and we have two observations for big data workloads:

First, \emph{Big data workloads have more branch instructions}: the average branch instruction percentage of big data workloads is 18.7\%, which is larger than those of HPCC, PARSEC, SPECFP and SPECINT distinctly. Furthermore, in big data workloads, from the application category dimension,
the average branch ratio of the service, data analysis and interactive analysis are 18\%, 19\% and 19\% respectively; from the system behavior dimension, the average branch ratio of the CPU-intensive, I/O intensive, and hybrid workloads are 19\%, 18\% and 19\% respectively.

The above observations can be explained as follows. 
First, big data analysis workloads (including data analysis and interactive analysis), as shown in Table 2, are based on discrete mathematics (such as graph computation), relation algebra and mathematical statistics (such as classification). This is different from the traditional numerical computation such as linear or differential equation in the scientific computing community. Traditional scientific computing code owns larger basic blocks as complex formula calculations \cite{murphy2007memory}, but big data analysis workload kernel code is biased to simple and conditional judgement operations. 
For example, Algorithm~\ref{Kmeans_algorithm} shows the kernel pseudocode of Kmeans, which includes a lot of judgements in the main loop (line 4 to 10) and the ComputeDist function block in the Algorithm~\ref{Kmeans_algorithm}  is simple, which contains only 40 lines in the real application; second, service workloads (Cloud OLTP) always need to conduct different processing steps to deal with diverse user requests, which result in many conditional judgment operations. The Switch-Case style is adopted frequently. This also is similar to the traditional service workloads, such as the TPC-C workload, which also has very high branch instruction ratio (30\%).

\begin{algorithm}[htbp]
\caption{Kmeans \textbf{\{\}}}
\label{Kmeans_algorithm}
\algsetup{
linenosize=\small,
linenodelimiter=.
}
\renewcommand{\algorithmicrequire}{\textbf{Input:}}
\renewcommand{\algorithmicensure}{\textbf{Output:}}
\begin{algorithmic}[1]
\REQUIRE Global variables $centers$;\\
The offset $key$;\\
The sample $value$.\\
\ENSURE $<key', value'>$ pair, where $key'$ is the index of the closest
center point and $value'$ is a string comprising of sample information.
\STATE Construct the sample $instance$ from $value$;
\STATE $minDis=Double.MAX\_VALUE$;
\STATE $index=-1$;
\FOR{$i=0$ to $centers.length$}
\STATE $dis=ComputeDist(instance, centers[i])$;
\IF{$dis<minDis$}
\STATE $minDis=dis$;
\STATE $index=i$;
\ENDIF
\ENDFOR
\STATE Take $index$ as $key'$;
\STATE Construct $value'$ as a string comprising of the values of
different dimensions;
\RETURN $<key', value'>$ pair;
\end{algorithmic}
\end{algorithm}

Furthermore, we profile the branch mis-prediction ratios on two X86 platforms: Intel Xeon E5645 and Intel Atom D510. Intel Atom D510 is a low-power processor with simple branch predictors and Intel Xeon E5645 is a server-processor with sophisticated predictors. Our result shows that the average branch mis-prediction ratio on Intel Atom D510 processor is 7.8\%, whereas the branch mis-prediction ratio on Intel E5645 processor
is only 2.8\%. As shown in Table \ref{Branch}, we found that E5645 has more sophisticated branch prediction mechanisms, e.g., the loop counter, indirect jumps and calls. Furthermore it is equipped with more BTB (Branch Target Buffer) entries.

\begin{table}[H]
\caption{The summary of branch prediction mechanisms.}\label{Branch} \center \begin{tabular}{|p{1.0in}|p{1.0in}|p{1.0in}|}
\hline
Component & D510 & E5645  \\ \hline
Conditional jumps&two-level adaptive predictor with a global history table &hybrid predictor combining a two-level predictor and a loop counter  \\ \hline
Indirect jumps and calls& Not &  two-level predictor \\ \hline
BTB Entries& 128& 8192 \\ \hline
Misprediction penalty & 15 cycles& 11-13 cycles \\ \hline
\end{tabular} \end{table}

Second, \emph{big data workloads have more integer instructions}: On Intel Xeon E5645, the average integer instruction ratio is 38\%, which is much higher than those of HPCC, PARSEC, SPECFP workloads. Also this value is close to that of CloudSuite (the ratio is 34\%) , SPECINT (the ratio is 41\%) and TPC-C(33\%) workloads, which are also integer dominated workloads. Furthermore, in big data workloads, from the application category dimension, the average integer instruction ratio of the service, data analysis and interactive analysis workloads are 40\%, 38\% and 38\% respectively; from the system behavior dimension, the average integer instruction ratio of the CPU-intensive, I/O intensive, and hybrid workloads are 37\%, 39\% and 38\% respectively.

The above observations can be explained as follows. First, many big data workloads are not floating-point dominated workloads (e.g. Sort, Grep, WordCount, and most of the Query and Cloud OLTP). Second, the floating-point dominated workloads such as Bayes, Kmeans and PageRank need to process massive amount of operations before they perform the floating-point operations. For example, the address calculation, branch calculation, all of which are integer operations.

\begin{figure}
\centering
\includegraphics[scale=0.7]{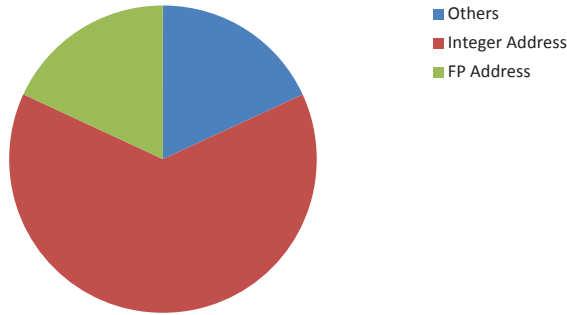}
\caption{The integer instruction breakdown on X86 platforms.}
\label{figure:InB}
\end{figure}

Furthermore, we analyze the integer instruction breakdown of big data algorithm through inserting the analysis code into the source code to analyze the details of integer operations. We classify all operations into three classes. The first class is integer address calculation, such as locating the position in the integer array; the second class is floating point address calculation, such as locating the position in the floating-point array; the third class is other calculations such as computations or branch calculations. From Figure~\ref{figure:InB}, we can see that the average integer instruction breakdown:  64\% is integer address calculating, 18\% is floating point address calculating and 18\% is other calculations in big data workloads.

Combining the results of Figure~\ref{figure:InsMix} and Figure~\ref{figure:InB}, the average ratio of load/store and address calculation
related instructions is rough 73\% for big data workloads. These instructions are all related with data movement. This ratio increases to 92\% when further considering the branch instructions. Hence it is reasonable to say that
big data workloads are data movement dominated computing with more branch operations.

In this subsection, we only investigate instruction mix based on the Intel Xeon E5645, which is a typical X86 pocessor. Blem etl. \cite{blem2013power} have proved that ISAs of RISC and CISC are irrelevant to modern microprocessor performance and the difference of instruction mix between different ISA sets is indistinguishable.

\textbf{[Implications]}.

High branch instruction ratio indicates that sophisticated branch prediction mechanisms should be used for big data workloads to reduce the branch mis-prediction, which will let pipeline flush all the wrong instructions
to fetch the correct ones and cause high penalty.

The large percentage of integer instruction implies that the floating point units in the processor should be designed appropriately to match the floating point performance. For examples, the E5645 processors can achieve 57.6 GFLOPS in theory, but the average floating point performance of big data workloads is about 0.1 GFLOPS. Furthermore the latest Xeon processor like  Dual Xeon E5 2697 can achieve 345 GFLOPS, thus incurring a serious waste of floating point capacity and hence die size.

\subsection{ILP}\label{ILP}
\begin{figure}
\centering
\includegraphics[scale=0.5]{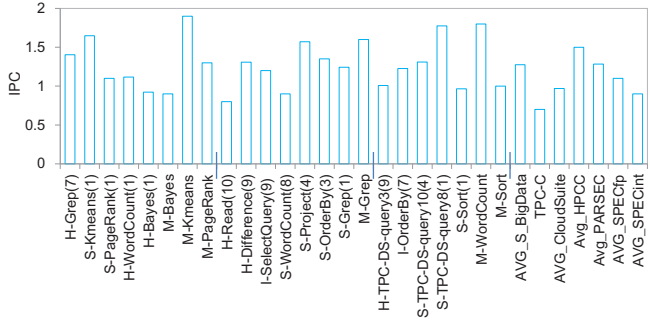}
\caption{The IPC of different workloads on the X86 platform.}
\label{figure:IPC}
\end{figure}

As shown in Figure~\ref{figure:IPC}, the average IPC of the big data workloads is 1.28. The average IPC of the big data workloads is larger than those of SPECFP (1.1) and SPECINT (0.9), as same as PARSEC (1.28), and slightly smaller than HPCC(1.5). This implies that the instruction-level parallelism of  big data workloads is not considerably different from other traditional workloads.

Furthermore, in big data workloads, from the application category dimension, the average IPC of the service, data analysis and interactive analysis workloads are 0.8, 1.2 and 1.3 respectively; from the system behavior, the average IPC of the CPU-intensive, I/O intensive, and hybrid workloads are 1.3, 1.2 and 1.3 respectively. So we can observe there are varied ILP for different category of applications. \emph{There are significant disparities of IPC among different big data workloads}, and the service workloads have lower IPC. This result corroborates with the report in~\cite{ferdman-asplos-clearing-the-clouds}, which confirmed most of services workloads in CloudSuite  have low IPC (the average number is 0.9). Similar to CloudSuite, we also notice that H-Read (0.8) is the only one service workload in the 17 workloads that has low IPC. However, a large percentage of the big data workloads (in BigDataBench subset) have higher IPC than the average number of SPECFP. Even several workloads have quite high IPC. Examples are S-Project (1.6) and S-TPC-DS-query8 (1.7). These observations have two points: first, CloudSuite only covers a part of the workloads included in BigDataBench in terms of IPC numbers. Second, there are significant disparities of IPC among different big data workloads.

\textbf{[Implications]}.  Architecture communities are exploring different technology road maps for
big data workloads: some focuses on scale-out wimpy core (e.g. in-order cores), for example,
HP's Moonshot uses Intel Atom processors ~\cite{Moonshot} and Facebook's interest is in ARM processors~\cite{Bill_Jia_BPOE_1};  others internet service providers try  to use brawny core or even accelerators, e. g., GPGPU for CPU-intensive computing like deep learning. Work in~\cite{ferdman-asplos-clearing-the-clouds} advocates use of modest degree of superscalar out-of-order execution. The ILP analysis of the big data workloads shows that there are different subclasses of big data workloads, which is also confirmed by our analysis of cache behaviors in Section~\ref{cache_behavior}. We speculate that the processor architecture should not have one-size-fits-all solution.

\subsection{Cache Behaviors}\label{cache_behavior}

\begin{figure}
\centering
\includegraphics[scale=0.5]{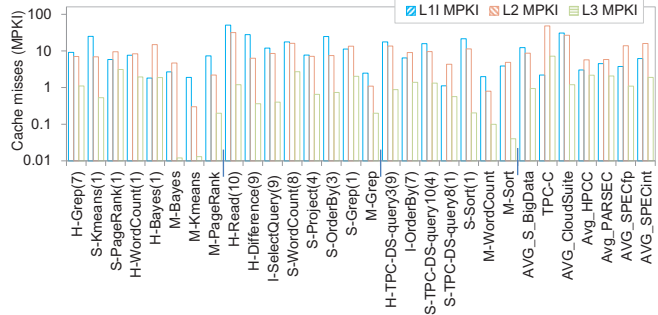}
\caption{The Cache behaviors of different workloads on the X86 platform.} \label{figure:Cache} \end{figure}

\textbf{L1I Cache Behaviors}:
As the front-end pipeline efficiencies are closely related with the L1I Cache behaviors, we evaluate the pipeline front-end pipeline efficiencies through investigating the L1I Cache MPKI (Miss per kilo instructions).
In Figure~\ref{figure:Cache}, the average L1I MPKI of the big data workloads is 15.  Corroborating previous work~\cite{ferdman-asplos-clearing-the-clouds}~\cite{jia2013characterizing}~\cite{hpca}, the average L1I MPKI of the big data workloads is larger than those of SPECFP, SPECINT, HPCC and PARSEC, but lower than that of CloudSuite (the average L1I is large as 32).

Furthermore, in big data workloads, from the application category dimension, the average L1I MPKI of the service, data analysis and interactive analysis workloads are 51, 13 and 14 respectively; from the system behavior dimension, the average L1I MPKI of the CPU-intensive, I/O intensive, and hybrid workloads are 8, 22, and 9, respectively. \emph{There  are significant disparities of L1I cache MPKI among different big data workloads}.
Among those workloads, H-Read has the highest L1I MPKI (51). The reason of high L1I MPKI for H-Read is that, as a service workload,
the user requests are more stochastic and  hence the instruction executions are more stochastic than other workloads.
So the instruction footprint should be more larger. We also investigated other service workloads.
Corroborating the report in~\cite{ferdman-asplos-clearing-the-clouds}, we found that most of them have high L1I MPKI.
Examples are TPC-C and Streaming, Olio, Cloud9, Search in the CloudSuite.
Our system behavior analysis in Section~\ref{Workloads} shows that most of the CloudSuite workloads will be classified into
the I/O intensive workloads according to our rule, and hence higher L1I MPKI is reported in CloudSuite.  These observations indicate  that CloudSuite only covers a part of the workloads included in the BigDataBench subset again.

\textbf{L2 Cache and LLC Behaviors}: As L1D cache miss penalty can be hidden by modern out-of-order pipeline, we evaluate the data-access efficiencies through investigating the L2 and L3 Cache MPKI.

In Figure~\ref{figure:Cache}, the average L2 MPKI of the big data workloads is 11.  The average L2 MPKI of the big data workloads is larger than those of HPCC and PARSEC, smaller than CloudSuite, TPC-C, SPECFP and SPECINT. Furthermore, in big data workloads, from the application category dimension, the average L2 MPKI of the service, data analysis and interactive analysis workloads are 32, 11 and 8 respectively; from the system behavior dimension, the average L2 MPKI of the CPU-intensive, I/O intensive, and hybrid workloads are 6.8, 11.6, and 7.7 respectively. \emph{There  are disparities of L2 cache MPKI among different big data workloads}. Among the workloads, The IO-intensive workloads and service workloads undergo more L2 MKPI.

In Figure~\ref{figure:Cache}, the average L3 MPKI of the big data workloads is 1.2.  The average L3 MPKI of the big data workloads is smaller than all of the other workloads. Furthermore, in big data workloads, from the application category dimension, the average L3 MPKI of the service, data analysis and interactive analysis workloads are 1.2, 1.7 and 0.8 respectively; from the system behavior dimension, the average L3 MPKI of the CPU-intensive, I/O intensive, and hybrid workloads are 1.7, 1.2, and 0.9 respectively. \emph{There are disparities of L3 cache MPKI among different big data workloads}. Among the workloads, The CPU-intensive workloads and data analysis workloads undergo more L3 MKPI.

\begin{figure}
\centering
\includegraphics[scale=0.5]{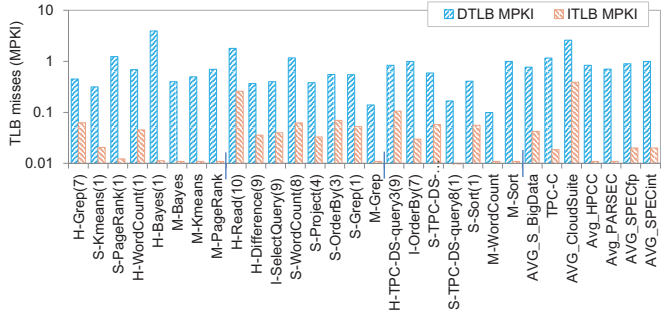}
\caption{The TLB behaviors of different workloads on X86 platforms.} \label{figure:TLB} \end{figure}

\textbf{TLB Behaviors}:
In Figure~\ref{figure:TLB}, the average ITLB MPKI of the big data workloads is 0.05. The average ITLB MPKI of the big data workloads is larger than those of HPCC, PARSEC, TPC-C, SPECFP and SPECINT, smaller than CloudSuite. Furthermore, from the application category dimension, the average ITLB MPKI of the service, data analysis and interactive analysis workloads are 0.2, 0.04 and 0.04 respectively; from the system behavior dimension, the average ITLB MPKI of the CPU-intensive, I/O-intensive, and hybrid workloads are 0.03, 0.08, and 0.05, respectively. The service and IO-intensive workloads undergo more ITLB MKPI.

The average DTLB MPKI of big data workloads is 0.9. The average DTLB MPKI of the big data workloads is close to those of HPCC and PARSEC, smaller than those of CloudSuite, TPC-C, SPECFP and SPECINT. Furthermore, from the application category dimension, the average DTLB MPKI of the service, data analysis and interactive analysis workloads are 1.8, 1.1 and 0.5 respectively; from the system behavior dimension, the average DTLB MPKI of the CPU-intensive, I/O-intensive, and hybrid workloads are 1.3, 0.7, and 0.6, respectively. The CPU-intensive and service workloads undergo more DTLB MKPI.

\textbf{[Implications]}
Our results show that in addition to varied instruction-level parallelism, there are significant disparities of front-end efficiencies among different big data workloads with respect to CloudSuite. These observations indicate that CloudSuite only covers a  part of the workloads included in BigDataBench.

\subsection{Locality}\label{Footprint}
As the cache behavior results on E5645 is interesting, such as the average L1I MPKI is higher, we further investigate it. The workload's instruction or data footprint can reflect the workload's locality, which is mainly related with the cache miss ratio metrics of the workload on the specific processor. In this subsection, we used simulator: MARSSx86, to evaluate the instruction and data footprint of the representative big data workloads. As the micro-architecture simulation is very time-consuming, we only choose the Hadoop workloads as the case study.

\subsubsection{Simulator Configurations}
The locality evaluation methods are similar to \cite{parsec} and \cite{2008parsec}. The temporal locality of a program can be estimated by analyzing the miss rate curve with the cache capacity changing. In this paper, we use MARSSx86 as the simulator and change the L1D and L1I cache size for evaluation.

\textbf{System Configurations}.

First, we set processor architecture as Atom-like in-order pipeline with a single core.
Second, we set  two-level cache with L1D, L1I and L2: 8-way associative L1 cache with 64 byte lines and shared 8-way associative L2 cache with 64 byte lines. Third, we change the L1 cache size from 16 KB to 8192 KB and record the cache miss ratio of each run.

\textbf{Workloads Configurations}.
For the Hadoop workloads, we choose the input data size as 64MB and set one map and one reduce slot. We use simsmall to drive PARSEC in the simulator. 

\textbf{Running Configurations}.
First, we performs the functional simulation: Qemu, to skip first billion of instructions which is the prepared stage of Hadoop, then we switch to detailed simulation: MARSSx86 to execute. We choose five segments of Hadoop workloads, which include Map 0\% to 1\%, Map50\% to 51\%, Map 99\% to 100\%, Reduce 0\% to 1\% and Reduce 99\% to 100\%. The cache miss ratio of hadoop workloads are the weighted mean of five segments.

\subsubsection{Experiments results}
\begin{figure}
\centering
\includegraphics[scale=0.8]{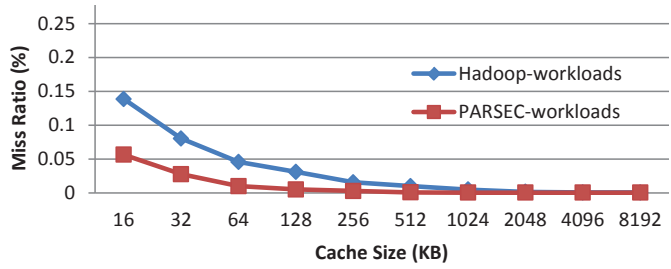}
\caption{Instructions Cache miss ratio versus Cache size.}
\label{figure:L1ITem}
\end{figure}

Figure~\ref{figure:L1ITem}  reports the average instruction cache miss ratios versus cache size for the representative big data workloads and PARSEC. From Figure~\ref{figure:L1ITem}, we can find that instruction cache miss ratio of the Hadoop workloads are larger than those of PARSEC workloads distinctly. The footprint of PARSEC is about 128 KB in experiments, but that of big data Hadoop workloads is about 1024 KB. The main reason should be that Hadoop has deep software stacks, which makes instruction footprint larger, and Jia ~\cite{jiacharacterization} also describes the same viewpoint from the code size.

\begin{figure}
\centering
\includegraphics[scale=0.8]{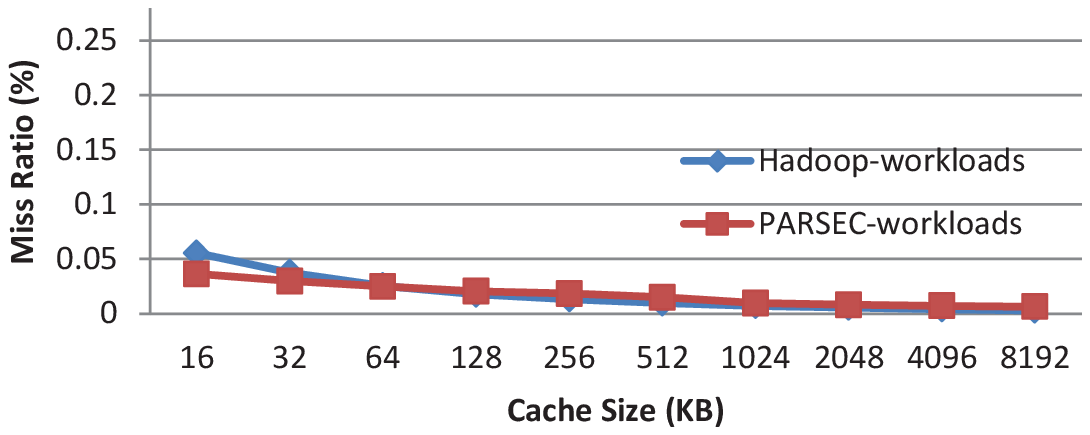}
\caption{Data Cache miss ratio versus Cache size.}
\label{figure:L1DTem}
\end{figure}

Figure~\ref{figure:L1DTem}  reports the average data cache miss ratios versus increasing cache size for the representative big data workloads and PARSEC. From Figure~\ref{figure:L1DTem}, we observe that the data cache miss ratio of PARSEC and Hadoop workloads are close after 64KB. This observation contradicts our intuition that  big data workloads should have larger data footprint as it processes huge data.

\begin{figure}
\centering
\includegraphics[scale=0.8]{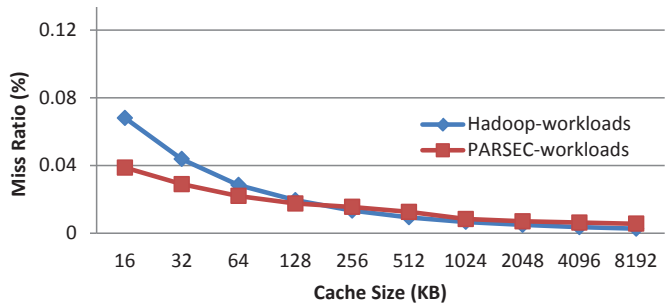}
\caption{Cache miss ratio versus Cache size.}
\label{figure:L1ATem}
\end{figure}

Figure~\ref{figure:L1ATem} reports the average cache miss ratios versus increasing cache size for the representative big data workloads and PARSEC. From Figure~\ref{figure:L1ATem}, we can see that the cache miss ratio of PARSEC and the  Hadoop workloads are close after 1024KB.

\textbf{[Implications]}.

The footprint should be more related with the cache efficiency of workloads on the processor. The footprint result is also in accord with the cache behaviors results. Furthermore, in the footprint experiments, we have two observations:  first, for the Hadoop workloads, which have complex software stacks, the capacity requirement of L1I cache capacity is more larger than that of traditional workloads; second, the capacity requirements of L1D cache and other level cache (such as L2 and L3) which are shared by instruction and data is not significantly different between the Hadoop workloads and the traditional workloads.

\subsection{Software Stacks Impacts}\label{SoftwareStacks}
\begin{figure}
\centering
\includegraphics[scale=0.8]{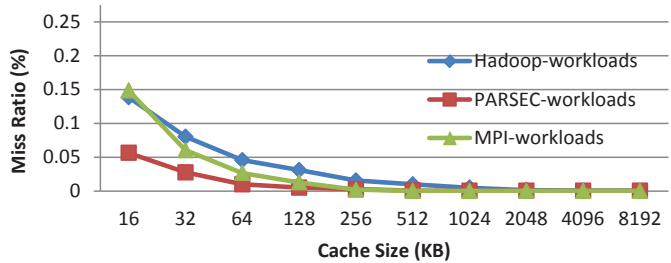}
\caption{Instructions Cache miss ratio versus Cache size.}
\label{figure:L1IMPI}
\end{figure}

In order to investigate the software stacks deeply, we also add six workloads implemented with MPI (the same workloads included in the representative big data workloads). As shown in from Figure.~\ref{figure:IPC} to Figure.~\ref{figure:L1IMPI}, we have the following observations:

First, for the same algorithms or operations, when implemented with different software stacks, the latter has a serious impact on IPC. As shown in Figure~\ref{figure:IPC}, the IPC of M-WordCount (implemented with MPI) is 1.8 while those of the Hadoop and Spark implementations are only 1.1 and 0.9, respectively. The average IPC of MPI workloads is 1.4 and that of the other workloads is 1.16. The gap is 21\%.

Second, for the same algorithms or operations, software stacks have significant impact on processor front-end behaviors. As shown in Figure~\ref{figure:Cache}, the MPI versions of the big data workloads have lower L1I MPKI. For example, the L1I MPKI of M-WordCount is 2 while those of the Hadoop and Spark implementations are 7 and 17, respectively, and there are one order of magnitude differences. Furthermore, we calculate the average L1I MPKI of the MPI version workloads, and the number  is only 3.4, while that of the Spark or Hadoop versions is 12.6. This implies that complex software stacks that fail to use state-of-practise processor efficiently are one of the main factors leading to high front-end stalls. Furthermore, we investigate the footprint of the MPI big data workloads. Figure~\ref{figure:L1IMPI}  shows the average instruction cache miss ratios versus cache capacity for the MPI-based big data workloads and PARSEC. From Figure~\ref{figure:L1IMPI}, we can see that instruction cache miss ratios of the MPI-version  big data workloads are equal to those of PARSEC workloads, and less than those of the Hadoop workloads. This implies that the instruction footprint of the MPI workloads are similar to those of PARSEC workloads.

Third, the software stacks significantly impact not only  the L1I behaviors but also the L2 cache and LLC behaviors. As shown in Figure~\ref{figure:Cache}, the L2 and L3 MPKI of M-WordCount are 0.8 and 0.1 while those of the Hadoop version are 8.4 and 1.9 respectively; those of the Spark version are 16 and 2.7, respectively.

\subsection {Summary}\label{Summary}
We summarize our observations as follows.

First, the big data workloads are data movement dominated computing with more branch operations, taking up to 92\% percentage in terms of instruction mix.

Second, there are significant disparities of IPC among different big data workloads. And the average instruction-level parallelism of the big data workloads on the  main-stream processor is not distinctively different from other traditional workloads.  We also note that there are significant disparities of front-end efficiencies among different big data workloads.

Third, complex software stacks that fail to use state-of-practise processor efficiently are one of the main factors leading to high front-end stalls. For the same workloads, the L1I cache miss rates have one order of magnitude differences among the diverse implementations using different software stacks.

Finally, we confirmed the observation in~\cite{jia_bigDataBench_subset}: software stacks have significant impact on others micro-architecture characteristics, e.g., IPC, L2 Cache, LLC behaviors. In addition to innovative hardware design, we should pay great attention to co-design of software and hardware so as to use state-of-practise processors efficiently.

\section{Related Work}

Big data attract great attention, appealing many research efforts on big data benchmarking.
Wang et al. \cite{hpca} develop a comprehensive big data benchmark suite--BigDataBench, but it consists of too many workloads resulting in expensive
overhead for conventional simulation-based methods. In comparison, Ferdman et al. \cite{ferdman-asplos-clearing-the-clouds} propose CloudSuite, consisting of seven scale-out data center applications, but they only select a few workloads according to the so-called popularity, leading to partial or biased observations, which is confirmed in Section~\ref{E5645}.
Xi et al. \cite{xi2011characterization} measure the microarchitectural characteristics of search engine.
Ren et al. \cite{ren2012workload} provide insight into performance and job characteristics via analyzing Hadoop traces derived from a 2000-node production Hadoop cluster in TaoBao.
Smullen et al. \cite{smullen2007benchmark} develop a benchmark suite for unstructured data processing, and present four benchmarks which capture data access patterns of core operations in a wide spectrum of unstructured data processing applications.
Huang et al. \cite{huang2010hibench} introduce HiBench, evaluating a specific big data platform---Hadoop with a set of Hadoop programs.

Much research work focuses on workload analysis. Jia et al. \cite{jia2013characterizing} characterize data analysis workloads in data centers. They conclude that inherent characteristics exist in many data analysis applications,
different from desktop, HPC, traditional server workloads, and scale-out service workloads.
Tang et al. \cite{tang2011impact} study the impact of sharing resources on five datacenter applications, including web search, bigtable, content analyzer, image stitcher and protocol buffer. They find that a sizable benefit and potential degradation exist from resource sharing effects.
Mishra et al. \cite{mishra2010towards} propose a task classification methodology, in consideration of workloads dimensions, clustering and break points of qualitative coordinates, and apply it to the Google Cloud Backend.
Jia et al. \cite{jia_bigDataBench_subset} evaluate the microarchitectural characteristics of big data workloads---BigDataBench. They find that software stacks e.g. MapReduce v.s. Spark have significant impact on user-observed performance and micro-architectural characteristics. Our work confirms this work again  such that we develop an open source workload characterization tool which can automatically reduce comprehensive workloads to a subset of representative workloads.

Several work proposes system-independent characterization approaches.
Hoste et al. \cite{hoste2007microarchitecture} measure the characteristics of 118 benchmarks by collecting microarchitecture-dependent and microarchitecture-independent characteristics.
Eeckhout et al. \cite{eeckhout2005exploiting,joshi2006measuring} exploit program microarchitecture independent characteristics and measure benchmark similarity.
Phansalkar et al. \cite{phansalkar2007subsetting} subset the SPEC CPU2006 benchmark suite in consideration of microarchitecture-dependent and microarchitecture-independent characteristics.
We will perform system-independent characterization work on representative big data workloads in near future.

\section{Conclusion}
In this paper, we choose 45 metrics from a perspective of micro-architectural characteristics. On the basis of a comprehensive big data benchmark suite---BigDataBench, we reduce 77 workloads to 17 representative one.

We compare the representative big data workload subset with SPECINT, SPECCFP, PARSEC, HPCC, CloudSuite, and TPC-C.   To investigate the impact of different software stacks, we also add six workloads implemented with MPI (the same workloads included in the representative big data workloads). We found that the big data workloads are data movement dominated computing with more branch operations, which takes up to 92\% percentage in terms of instruction mix. Comparing with the traditional workloads i. e. PARSEC, the big data workloads have larger instruction footprint. Furthermore, there are significant disparities of front-end efficiencies among different subclasses of big data workloads. Finally, software stacks that fail to use state-of-practise processors efficiently are one of the main factors leading to high front-end stalls. In addition to innovative hardware design, we should pay great attention to co-design of software and hardware so as to use state-of-practise processors efficiently.

\parskip=0pt
\parsep=0pt
\bibliographystyle{ieeetrsrt}
\bibliography{myref}


\end{document}